\documentclass[final]{aipproc}
\layoutstyle{8x11single}
\usepackage{color}
\usepackage{amsmath}
\usepackage{graphics}
\usepackage{graphicx}

\newcommand{\be}{\begin{equation}}
\newcommand{\ee}{\end{equation}}
\newcommand{\ba}{\begin{eqnarray}}
\newcommand{\ea}{\end{eqnarray}}
\newcommand{\ar}{\arrowvert}

\newcommand{\bw}{\begin{widetext}}
\newcommand{\ew}{\end{widetext}}

\newcommand{\Imag}{\mathop{\mathrm{Im}}}
\newcommand{\Real}{\mathop{\mathrm{Re}}}

\begin{document}
\title{Strongly Interacting Electroweak Symmetry Breaking Sector with a Higgs-like light scalar}


\author{Antonio Dobado (speaker)}{
  address={Departamento de F\'{\i}sica Te\'orica I, Universidad Complutense de Madrid, 28040 Madrid, Spain}
}

\author{Rafael L. Delgado}{
  address={Departamento de F\'{\i}sica Te\'orica I, Universidad Complutense de Madrid, 28040 Madrid, Spain}
}

\author{Felipe J. Llanes-Estrada}{
  address={Departamento de F\'{\i}sica Te\'orica I, Universidad Complutense de Madrid, 28040 Madrid, Spain}
}

\begin{abstract}
The apparent finding of a 125-GeV light Higgs boson closes unitarity of the minimal Standard Model (SM), that is weakly interacting: this is an exceptional feature not generally true if new physics exists beyond the mass gap found at the LHC up to 700 GeV. 
Such new physics induces departures of the low-energy dynamics for the minimal electroweak symmetry-breaking sector, with three Goldstone bosons (related to longitudinal W bosons) and one light scalar, from the SM couplings. We calculate the scattering amplitudes among these four particles and their partial-wave projections in effective theory. For this we employ the Electroweak Chiral Lagrangian extended by one light scalar and carry out the complete one-loop computation at high energy including the counterterms needed for perturbative renormalization, of dimension eight. 
For most of parameter space, the scattering is strongly interacting (with the SM a remarkable exception). We therefore explore various unitarization methods, that can already be applied to the tree-level $W_L W_L$ amplitude; we find and study a natural second sigma-like scalar pole there.
\end{abstract}

\maketitle

\section{Introduction: the new experimental $S(125)$}
Numerous papers are addressing the issue of strong interactions in the electroweak sector of the Standard Model; this work leans on~\cite{Delgado:2013hxa,Delgado:2013loa}. 
The renaissance of strong beyond the Standard-Model physics (BSM) comes about 
because the LHC experiments ATLAS~\cite{ATLAS} and CMS~\cite{CMS} in Run-I have apparently found~\cite{twophotons} a SM Higgs-like boson (that is, a boson reported to have scalar quantum numbers and couplings compatible with those of a Standard-Model Higgs) but no new particle~\cite{searches} up to an energy of about 600-700~GeV (and higher yet for new vector bosons). The mass of the new boson is reported to be $M_\varphi\simeq 125$ GeV, of the same magnitude as the gauge boson masses $M_W$ and $M_Z$, leaving a mass gap in the spectrum between $M_\varphi$ and 700 GeV.

The presence of such a mass gap naturally suggests that the SM Higgs--like boson is an additional Goldstone boson, the other ones being the longitudinal components of gauge boson pairs $W_L W_L$ and $Z_L Z_L$. There are several models of new physics which support such an idea, like dilaton models (spontaneous breaking of scale invariance), or a composite Higgs based on $SO(5)/SO(4)$ or other cosets.

In this work as in~\cite{Delgado:2013hxa,Delgado:2013loa}, we explore the most general effective Lagrangian for $\sqrt{s}\gg M_W, \ M_Z,\ M_\varphi$ with the experimentally known particles which includes those models (and the SM) as particular cases. 

Our only separation from full generality will be a counting ansatz: we assume that the Higgs-potential self-couplings are of order $M_\varphi^2$, and thus negligible for $s\gg M_\varphi^2$. This assumption is shared by all models of interest at the present time (and, in particular, those mentioned here), and is our only assumption about the nature of the new physics.

The spectra of the longitudinal components of gauge-boson pairs $W_L W_L$  and $Z_L Z_L$ are not yet available, but they are expected to be measured in the next few years, when the LHC runs at 14~TeV.

The hypothetical presence of strong interactions makes electroweak perturbation theory less useful, but the equivalence theorem~\cite{ET} still applies to the scattering of longitudinal bosons $W_L W_L$ and $Z_L Z_L$ at high energy (high in the sense that the energy is high when compared with $M_W$ and $M_H=M_\varphi$, but not larger than about $4\pi v\simeq 3\,{\rm TeV}$).
The equivalence theorem allows us to substitute the $W_LW_L$ scattering amplitude for the Goldstone boson-Goldstone boson one with a controlled error,
\be
T(\omega^a\omega^b \rightarrow \omega^c\omega^d )  
= T(W_L^aW_L^b \rightarrow W_L^cW_L^d ) + O\left(\frac{M_W}{\sqrt{s}}\right)\ . 
\ee
This theorem applies to any renormalizable gauge and, in particular, to the Landau gauge which is very convenient when dealing with the Higgs mechanism.

\section{Generic electroweak effective Lagrangian and chiral formulation}

The general Lagrangian for the electroweak SBS at low energies, describing the dynamics of the four light modes (three $\omega$ WBGB and the Higgs $\varphi$), can be written~\cite{Delgado:2013loa} as
\be \label{genericLagrangian}
{\cal L}=\frac{v^2}{4}g(\varphi/f)Tr(D_\mu U)^\dag
D^\mu U+\frac{1}{2}\partial_\mu \varphi \partial^\mu
\varphi-V(\varphi) \ee
where $g(\varphi/f)$ is, in a general theoretical framework, an arbitrary analytical function,
\be
g(\varphi / f) = 1 + \sum_{n=1}^\infty g_n\left(\frac{\varphi}{f}\right)^n = 1 + 2\alpha\frac{\varphi}{f} + \beta\left(\frac{\varphi}{f}\right)^2 + \cdots,
\ee
(the often used $a$ and $b$ parameters~\cite{scalar} would be $a = \alpha v / f$ and $b=\beta v^2 / f^2$). 
Here, $U$ is a field on $SU(2)$, with parametrization chosen as $U = \sqrt{1 - \tilde{\omega}^2/v^2} + i\tilde{\omega}/v$ ($\tilde{\omega}=\omega_a\tau^a$). The gauge-covariant derivative is $D_\mu U = \partial_\mu U + W_\mu U - U Y_\mu$, the gauge fields being $W_\mu = -g i W_\mu^i \tau^i/2$ and $Y_\mu = -g' i B_\mu^i \tau^3/2$, inducing the coupling with transverse modes (which in this work will be neglected, g = g' = 0). The electroweak-breaking scale is as usual $v = 1/(\sqrt{2}G_F)\simeq 246 GeV$ and $f$ is a new arbitrary dynamical energy scale.

Because no two--Higgs final state has been observed at the LHC, there is no relevant (order O(1)) constraint on $\beta$. However, the available $WW$ data from CMS and ATLAS~\cite{Aad:2013wqa,CMS:aya} does constrain $\alpha/f$; at the 2-$\sigma$ confidence level,
\begin{eqnarray} \label{boundexp1}
f/\alpha &\in & (225,350){\rm GeV}  \ \ {\rm or}\  \  a \in (0.70,1.1) \ \ {\rm (CMS)} \\ 
\label{boundexp2}
f/\alpha&\in & (185,285){\rm GeV} \ \ {\rm or}\  \ a \in (0.87,1.3) \ \ {\rm (ATLAS)}\ .
\end{eqnarray}
Further (theory-dependent) bounds can be found in~\cite{bounds}.

The effective Lagrangian in Eq.~(\ref{genericLagrangian}), when simplified (and assuming $g=g'=0$), becomes
\begin{eqnarray} \label{bosonLagrangian_tree}
{\cal L} &=& \frac{1}{2}g(\varphi/f) \partial_\mu \omega^a
\partial^\mu \omega^b\left(\delta_{ab}+\frac{\omega^a\omega^b}{v^2-\omega^2}\right)   
\nonumber +\frac{1}{2}\partial_\mu \varphi \partial^\mu \varphi \nonumber \\
&&-\frac{1}{2}M^2_\varphi \varphi^2-\lambda_3 \varphi^3- \lambda_4 \varphi^4+...
\end{eqnarray}

Particular cases would be the SM-Higgs ($f=v$, $\alpha=\beta=1$, $g_i=0\ \ \forall i\ \geq 3$, $\varphi\rightarrow H$), dilaton models ($f\neq v$, $\alpha=\beta=1$, $g_i=0\ \ \forall \ i\geq 3$) or $SO(5)/SO(4)$ Minimal Composite Models with $\alpha= \cos \theta /\sqrt \xi, \beta = \cos(2\theta)/\xi$ and $\sin \theta = \sqrt \xi $
($\xi :=v^2 /f^2$).

This Lagrangian provides us with the LO tree-level and with the loop part of the NLO $\omega\omega\to \omega\omega$ scattering amplitudes, as well as 
those of the unitarity-related processes  $\omega\omega \rightarrow \varphi\varphi$ and $\varphi\varphi  \rightarrow \varphi\varphi$.
Near the the chiral limit (or equivalently at high energy), all the terms in the second line of Eq.~(\ref{bosonLagrangian_tree}) are negligible.

At NLO, the most general Lagrangian density has additional tree-level counterterms with four derivatives;
 provided that $s\gg M_W^2 \sim M_H^2$ (chiral limit), the minimum ones that carry out the NLO renormalization are~\cite{Delgado:2013hxa}
\begin{eqnarray} \label{higherorderL}
{\cal L}_4 & = &  a_4(tr V_\mu V_\nu)^2 +  a_5(tr V_\mu V^\mu)^2 \\ \nonumber
 & + &\frac{\gamma}{f^4} (\partial_\mu\varphi \partial^\mu\varphi)^2
 +  \frac{\delta}{f^2} (\partial_\mu\varphi \partial^\mu\varphi)tr(D_\nu U)^\dag D^\nu U
+\frac{\eta}{f^2} (\partial_\mu\varphi \partial^\nu\varphi)tr(D_\nu U)^\dag D^\mu U+...
\end{eqnarray}
where $V_\mu= D_\mu U U^\dagger$. These terms should be added to the Lagrangian density in Eq.~(\ref{genericLagrangian}). In this high energy $s\gg M_W^2 \sim M_H^2$ limit, simplifying Eq.~(\ref{genericLagrangian}) plus the counterterms in Eq.~(\ref{higherorderL}), we recover the one loop-renormalized effective theory described by
\ba \label{bosonLagrangian_loop}
{\cal L} & = & \frac{1}{2}\left(1 +2 \alpha \frac{\varphi}{f} +\beta\left(\frac{\varphi}{f}\right)^2\right)
\partial_\mu \omega^a
\partial^\mu \omega^b\left(\delta_{ab}+\frac{\omega^a\omega^b}{v^2}\right)   
\nonumber +\frac{1}{2}\partial_\mu \varphi \partial^\mu \varphi  \nonumber  \\
 & + & \frac{4 a_4}{v^4}\partial_\mu \omega^a\partial_\nu \omega^a\partial^\mu \omega^b\partial^\nu \omega^b +
\frac{4 a_5}{v^4}\partial_\mu \omega^a\partial^\mu \omega^a\partial_\nu \omega^b\partial^\nu \omega^b  +\frac{\gamma}{f^4} (\partial_\mu\varphi \partial^\mu\varphi)^2  \nonumber   \\
 & + & \frac{2\delta}{v^2f^2} \partial_\mu\varphi \partial^\mu\varphi\partial_\nu \omega^a  \partial^\nu\omega^a
+\frac{2\eta}{v^2f^2} \partial_\mu\varphi \partial^\nu\varphi\partial_\nu \omega^a \partial^\mu\omega^a.
\ea

\section{Unitarity violation in perturbation theory and various unitarization methods}
The unitarity condition for the exact reaction matrix $\tilde{T}$, 
reduces, for massless particles (as applicable when $s\gg M_W^2$), to
\be \label{CondicionUnitariedad}
  \Imag \tilde T = \tilde T \tilde T^\dagger\ .
\ee 
To implement unitarity it is very useful to introduce the partial waves:
 \be
 t_{IJ}(s)=\frac{1}{64 \pi}\int _{-1}^1 d(\cos  \theta)P_J(\cos \theta)  T_I(s,\cos \theta),
 \ee
where we are denoting by $ P_J(\cos \theta) $ the Legendre polynomials; $\theta$, the scattering angle in the center of mass frame; and $T_I$, the (custodial) isospin amplitudes.
At tree level (and in the region $M_\varphi^2\ll s$), the  Lagrangian in Eq.~(\ref{bosonLagrangian_tree}) 
yields the lowest, scalar ($l=J=0$) and weak-isoscalar, partial waves for the scattering processes $\omega\omega\rightarrow\omega\omega$, $\omega\omega\rightarrow\varphi\varphi$, $\varphi\varphi\rightarrow\omega\omega$ and $\varphi\varphi\rightarrow\varphi\varphi$:
\begin{eqnarray} \label{treewaves}
t_\omega(s) &=& 
\frac{1}{16\pi v^2}\left[ s(1-\xi \alpha^2)-\xi \alpha^2 M_\varphi^2 +
\xi  \alpha^2 \frac{M_\varphi ^4}{s}\log\left(\frac{s}{M_\varphi^2}\right)\right] \\
t_{\omega\varphi}(s) &=&
t_{\varphi \omega}(s)=
\frac{\sqrt{3}(\alpha^2-\beta)s}{32\pi f^2}+
\frac{\sqrt{3}\alpha^2}{16\pi f^2}\left[M_\varphi^2+ \frac{2 M_\varphi
^4}{s}\log\left(\frac{s}{M_\varphi^2}\right)\right]  \\
t_{\varphi}(s) &=&  
\frac{9
\lambda_3^2}{4 \pi s }  \log\left(\frac{s}{ M_\varphi^2}\right)-\frac{3\lambda_4}{4\pi} 
\end{eqnarray}

where the simplified notation $\omega = 2\omega 2\omega$, $\omega\varphi=2\omega2\varphi$ and $2\varphi 2 \varphi= \varphi$ has been used.  Of course the SM is perturbatively unitary, which is a very good approximation since the model is weakly interacting. However, when  $a \equiv \xi\alpha^2\neq 1$, the tree level partial wave $t_\omega(s)$ grows linearly at high $s$. That is, we deal with \emph{strong interactions}. Always remembering our kinematic region $s\gg M_\varphi^2$, the dominant terms at low energies become:
\begin{eqnarray} \label{LETS}
t_\omega  & \to & \frac{s}{16 \pi v^2}(1-\xi \alpha^2)   \nonumber  \\
 t_{\omega\varphi}  & \to  & \frac{\sqrt{3}(\alpha^2-\beta)s}{32 \pi f^2}   \nonumber  \\
 t_\varphi  & \to & 0
\end{eqnarray}
As unitarity is being broken in the presence of beyond SM physics, we need unitarization methods to explore the region of $\sim 1\,{\rm TeV}$. For the tree level computations, the $K$ matrix, large $N$, $N/D$ and IAM methods have been considered and all the unitarized expressions can be found in ref.~\cite{Delgado:2013loa}. As an example, let us quote the case of the $K$ matrix method. The unitarized amplitude matrix $\tilde{T}$ would be, in general,
\be
\tilde{T} = T(1-J(s)T)^{-1},\ \ \ \ J(s) = -\frac{1}{\pi}\log\left[\frac{-s}{\Lambda^2}\right]
\ee
And, in particular, the unitarized $\omega\omega$ elastic scattering amplitude $\tilde{t}_\omega$ would be
\be
\tilde{t}_\omega = \frac{t_\omega - J(t_\omega t_\varphi - t^2_{\omega\varphi})}{1-J(t_\omega +t_\varphi)+J^2(t_\omega t_\varphi - t^2_{\omega\varphi})},
\ee
which, for $\beta=\alpha^2$ (elastic case), would simplify to
\be
\tilde{t}_\omega = \frac{t_\omega}{1-J t_\omega}
\ee

\begin{figure}
\null\includegraphics[width=.45\textwidth]{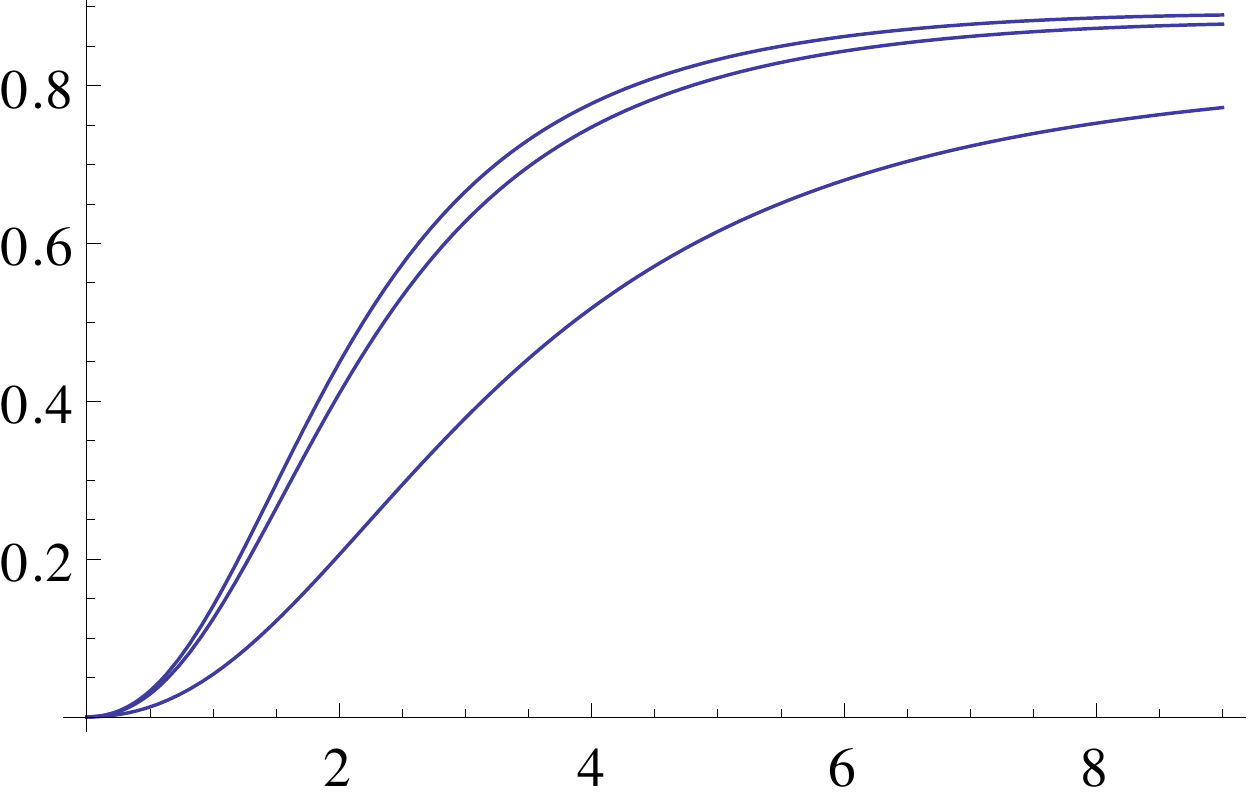}\hspace{.1\textwidth} %
\includegraphics[width=.45\textwidth]{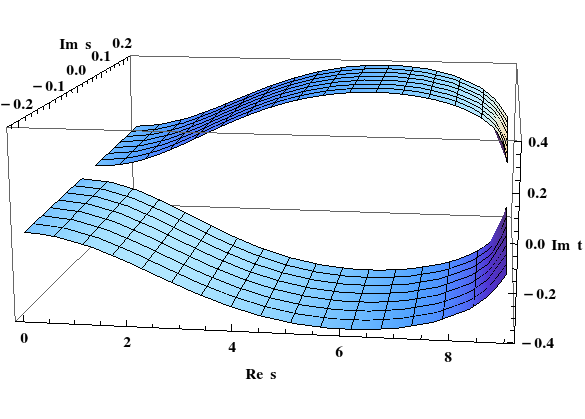}\null
\caption{On the left, $\lvert \tilde{t}_\omega\rvert$ vs. $s$ (TeV$^2$), unitarized with the K matrix method. $\Lambda=3\,{\rm TeV}$, $\mu=100\,{\rm GeV}$. From top to bottom, the lines are $f=1.2,\,0.8,\,0.4\,{\rm TeV}$. On the right, $\Imag\tilde{t}_\omega$ vs. $s$ (TeV$^2$), unitarized with the N/D method. $f = 1\,{\rm TeV}$, $\beta=1$, $m=150\,{\rm GeV}$.}\label{graf_K_ND}
\end{figure}

\begin{figure}
\null\includegraphics[width=.45\textwidth]{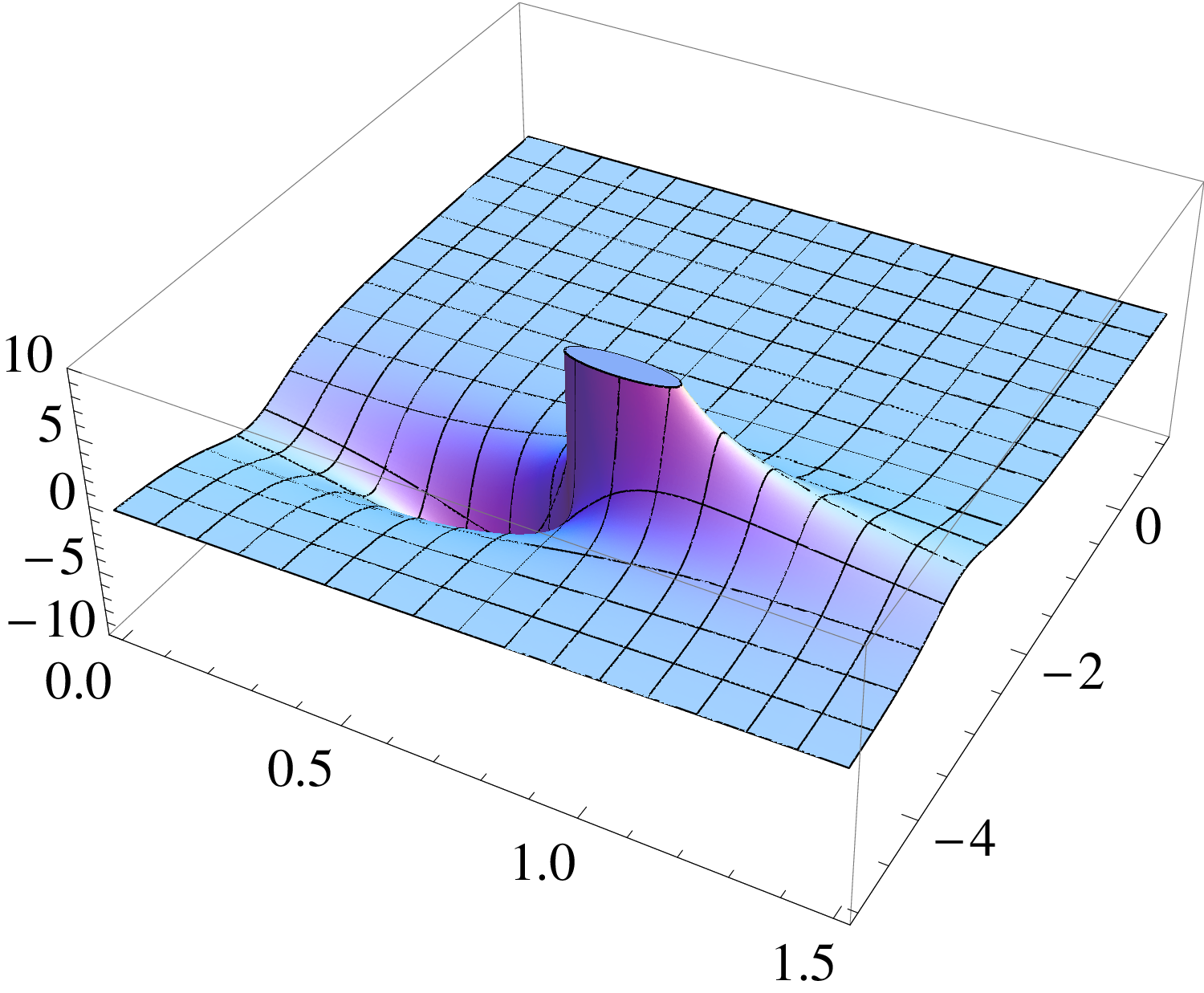}\hspace{.1\textwidth} %
\includegraphics[width=.45\textwidth]{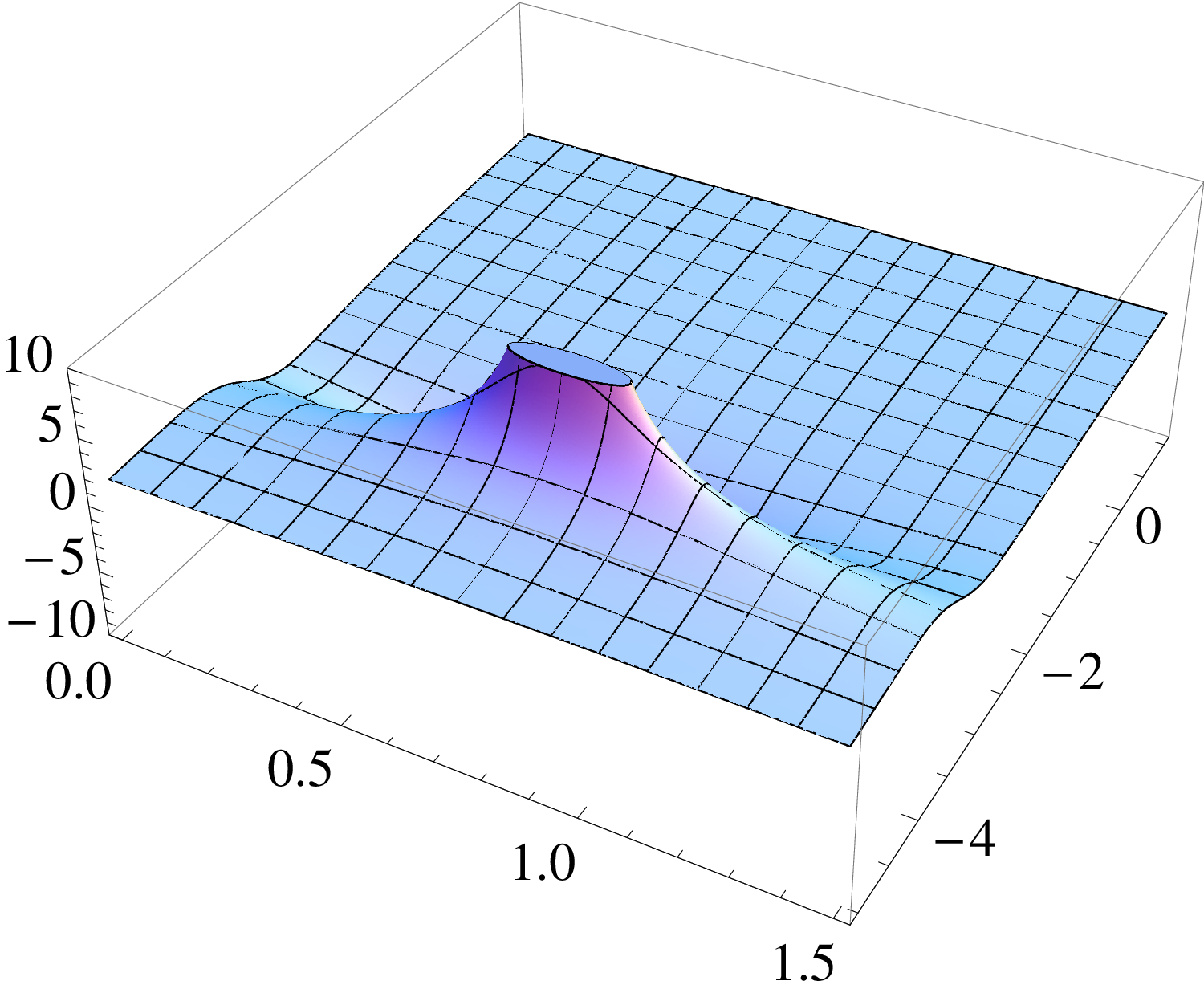}\null
\caption{Pole of the scalar $\omega\omega$ amplitude in the second Riemann sheet. $\Real\tilde{t}_\omega$ (left) and $\Imag\tilde{t}_\omega$ (right), unitarized with the large $N$ method, for $f = 400\,{\rm GeV}$. On the OX axis, $s$ in TeV$^2$.}\label{graf_largeN}
\end{figure}

\begin{figure}
\null\hfill\includegraphics[width=.8\textwidth]{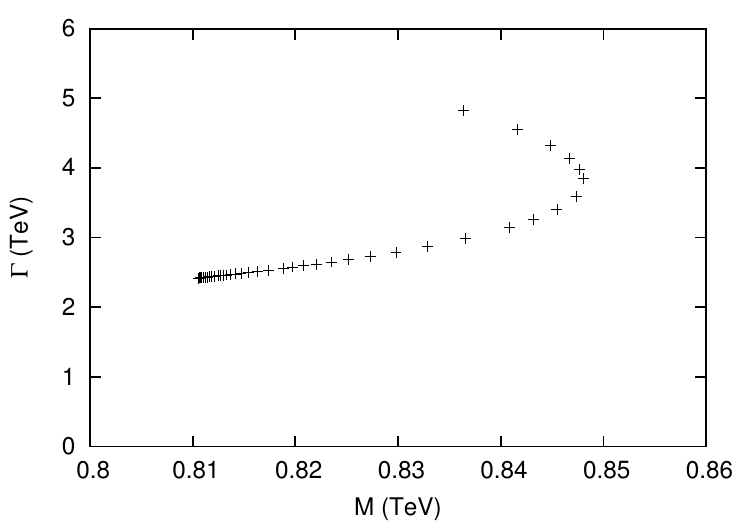}\hfill\null
\caption{Motion of the pole position of $\lvert\tilde{t}_\omega\rvert$, unitarized with the K--matrix method. $M_\varphi=125\,{\rm GeV}$, $f\in (250\,{\rm GeV},\,6\,{\rm TeV})$. The highest values of $f$ corresponds to the points near $M=810\,{\rm GeV}$.}\label{graf_pole_motion}
\end{figure}

Analytically continuing the amplitude to complex Mandelstam-s, a pole on the second Riemann sheet appears, as is shown in graphs~\ref{graf_K_ND} and~\ref{graf_largeN}. In graph~\ref{graf_pole_motion}, we have analyzed the motion of such pole upon varying the BSM scale $f$.

\begin{figure}
\null\includegraphics[width=.45\textwidth]{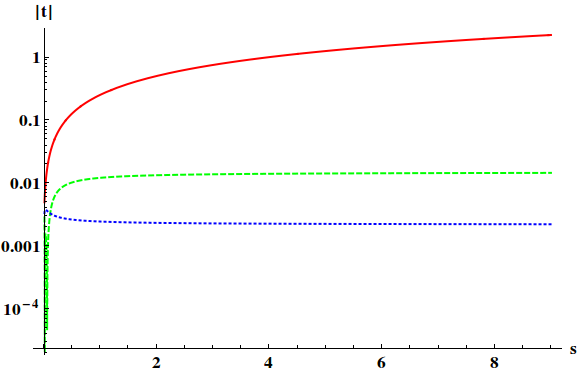}\hspace{.1\textwidth} %
\includegraphics[width=.45\textwidth]{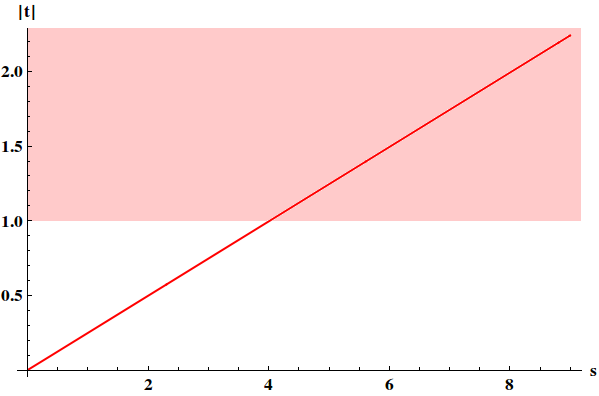}\null
\caption{On the left, $\lvert t\rvert$ (at tree level, eq.~\ref{treewaves}) vs. $s$ (TeV$^2$) for $f=2v$, $\beta=\alpha^2=1$, $\lambda_3=M^2_\varphi/f$, $\lambda_4=M^2_\varphi/f^2$. As seen, the amplitude for $\omega\omega$ scattering (solid red line) saturates unitarity, and is much larger than that for the  $\varphi\varphi$ channel (dashed green line) and the interchannel coupling amplitude is smallest (dotted blue line). On the right, $\lvert t_\omega\rvert$ vs. $s$ (TeV$^2$), for $f\ne v$ and without assuming $M_\varphi^2\ll s$. The violation of unitarity can be clearly seen. (The shaded area is unphysical.)}
\end{figure}

\section{One loop amplitudes}
We have also computed the one loop amplitudes from the Lagrangian in Eq.~(\ref{bosonLagrangian_loop}), using dimensional regularization $D=4-\epsilon$. We 
have checked our results in different limits and also  that our results agree with those found in \cite{Espriu:2013fia} in the limit of vanishining light scalar mass.

The complete expressions can be seen in ref.~\cite{Delgado:2013hxa}. For the sake of brevity, only the elastic $\omega\omega$ scattering amplitudes will be shown here. Such amplitudes $\omega_a\omega_b \rightarrow \omega_c\omega_d$ with isospin indices $a$, $b$, $c$, $d$, are best organized as
\be
{\mathcal T_\omega}_{abcd}= T_\omega(s,t,u)\delta_{ab}\delta_{cd}+T_\omega(t,s,u)\delta_{ac}\delta_{bd}+
T_\omega(u,t,s)\delta_{ad}\delta_{bc},
\ee 
where
\be \label{loopexpansion}
T_\omega = T_\omega^{(0)} + T_\omega^{(1)} \dots =  T_\omega^{(0)} + T^{(1)}_{\omega\ \rm tree} + T^{(1)}_{\omega \ \rm loop} \dots
\ee

The tree-level amplitudes $T_\omega^{(0)}$ and $T^{(1)}_{\omega\  \rm tree}$ are
\begin{equation} \label{Atree}
T_\omega^{(0)}(s,t,u) + T^{(1)}_{\omega\ \rm tree}(s,t,u) = (1-\alpha^2\xi)\frac{s}{v^2} + \frac{4}{v^4}\left[2a_5 s^2 + a_4(t^2 + u^2)\right].
\end{equation}
And the one-loop amplitude $T^{(1)}_{\omega  \ \rm loop}$,
\begin{equation} \label{Aloop}
T^{(1)}_{\omega \ \rm loop}(s,t,u) = \frac{1}{36 (4\pi)^2 v^4}[f(s,t,u)s^2 +(\alpha^2\xi-1)^2( g(s,t,u) t^2 + g(s,u,t) u^2)],
\end{equation}
where the following auxiliary functions have been used
\begin{eqnarray}
f(s,t,u) &:=& 
[20 - 40 \alpha^2\xi + \xi^2(56 \alpha^4 - 72 \alpha^2 \beta + 36 \beta^2)] \\
& + & [12 - 24 \alpha^2\xi + \xi^2(30 \alpha^4- 36 \alpha^2 \beta + 18 \beta^2)] N_\varepsilon  \\  \label{ffunction}
& + & [-18 + 36 \alpha^2\xi +\xi^2(- 36 \alpha^4 + 36 \alpha^2 \beta - 18 \beta^2)] \log\left(\frac{-s}{\mu^2}\right) \\
& + & 3 (\alpha^2\xi-1)^2 \left[\log\left(\frac{-t}{\mu^2}\right) +
\log\left(\frac{-u}{\mu^2}\right)\right] \\ \label{gfunction}
g(s,t,u) &:=& 
26 + 12 N_\varepsilon 
-9 \log\left[-\frac{t}{\mu^2}\right]
-3 \log\left[-\frac{u}{\mu^2}\right]
\end{eqnarray}
and in dimensional regularization $D=4-\epsilon$ the poles are contained as usual in
\be
N_\epsilon =\frac{2}{\epsilon} + \log 4\pi -\gamma\ .
\ee
These divergent terms can be removed by a proper renormalization of the $a_4$ and $a_5$ parameters in the elastic case and $\gamma, \delta$ and $\eta$ for the rest of the channels including the scalar field $\varphi$.

\begin{figure}
\null\includegraphics[width=.5\textwidth]{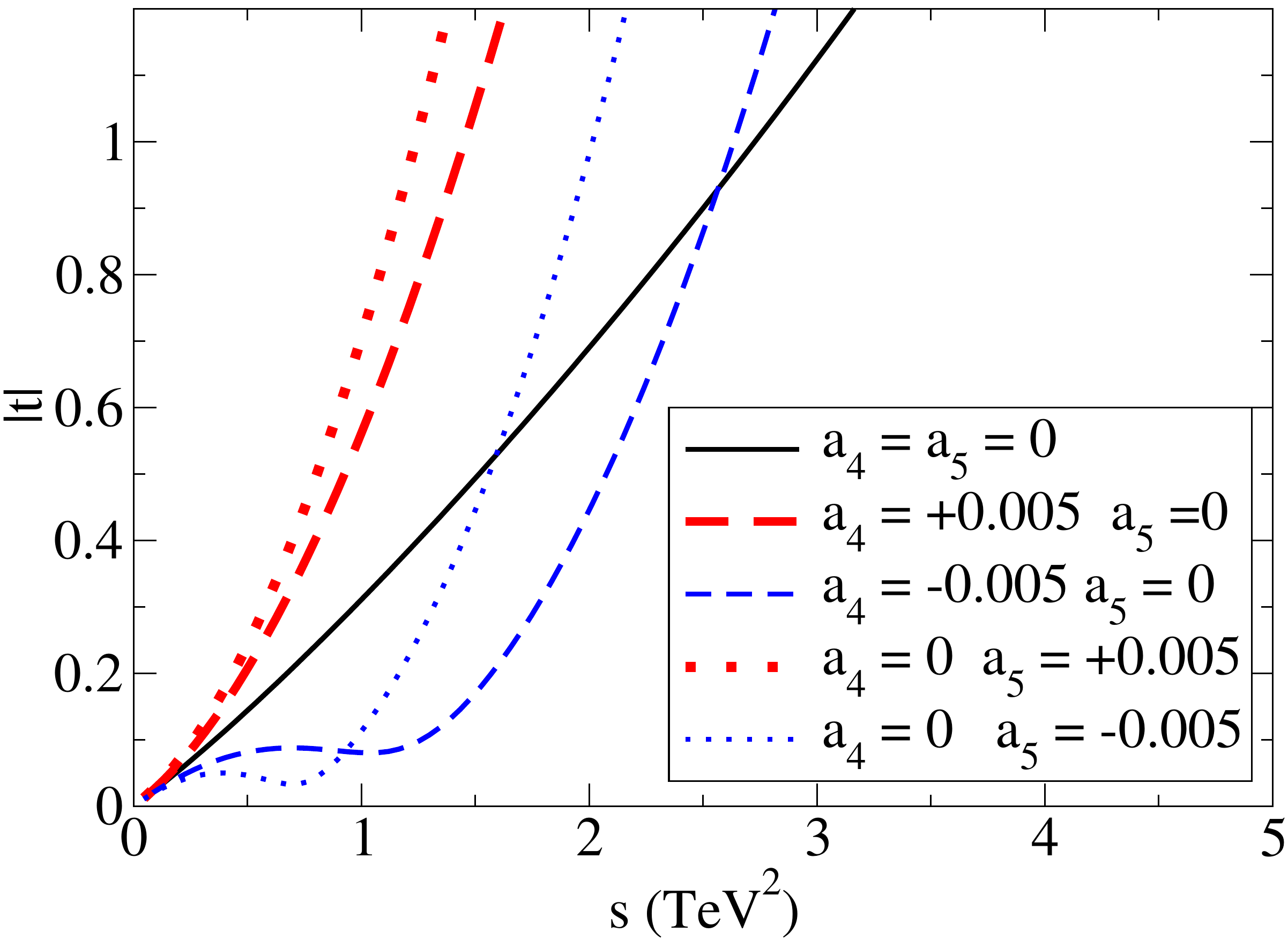}
\includegraphics[width=.5\textwidth]{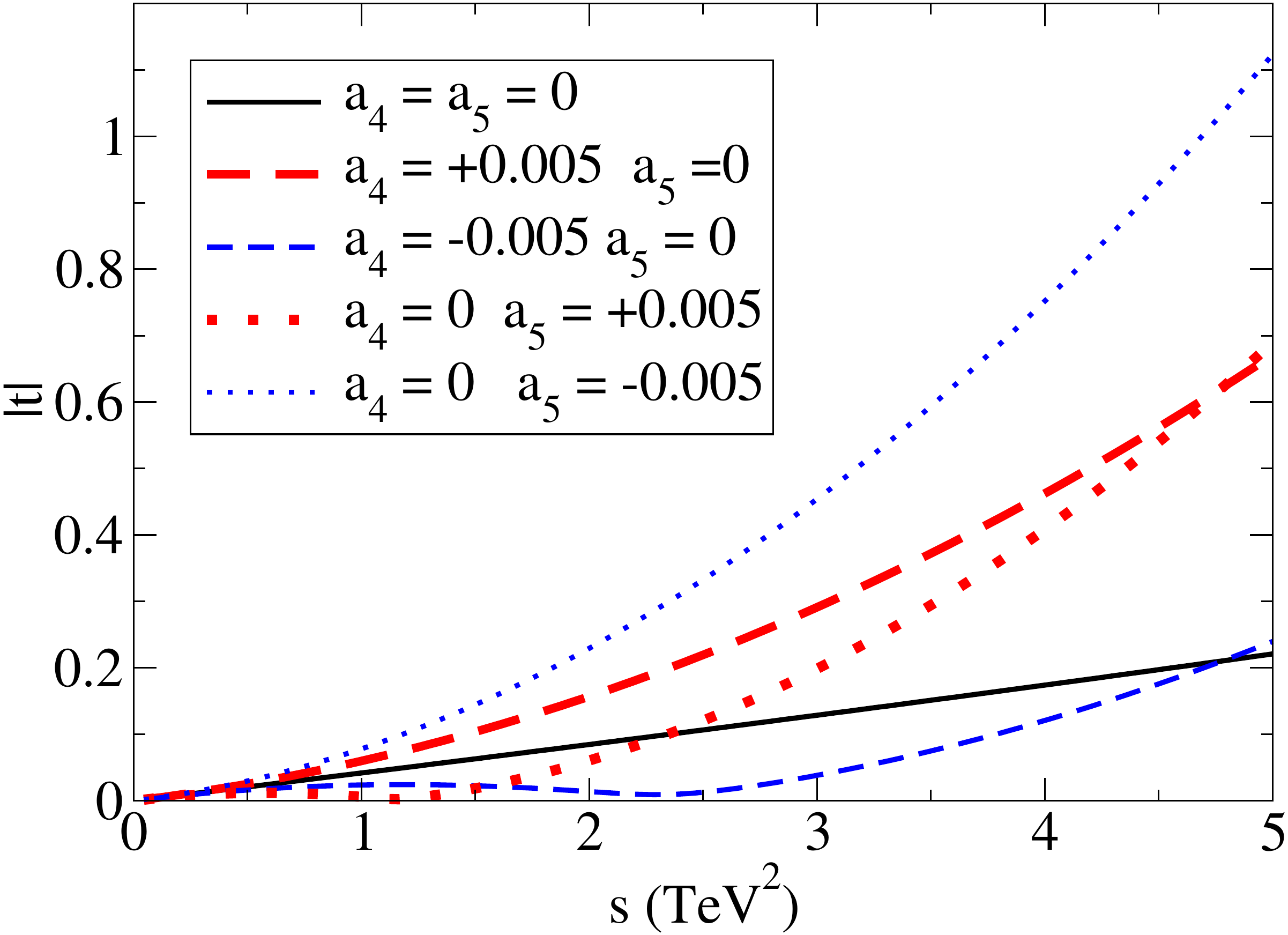}\null
\caption{One-loop elastic $\omega\omega\to\omega\omega$ partial wave amplitudes $t_{\omega\ IJ}$ for $f=500$ GeV and $\mu=1$ TeV. We take a non-zero (positive or negative) $a_4$ or $a_5$. From left to right, $\ar t\ar_{\omega\ 00}$ ($I=J=0$) and $\ar t\ar_{\omega \ 11}$ ($I=J=1$).\label{oneloopfig}}
\end{figure}

Figure~\ref{oneloopfig} shows two example (the projected scalar-isoscalar and the vector-isovector) perturbative amplitudes.

\section{Discussion and outlook}
Numerous authors have recently studied or revisited effective Lagrangians including a light Higgs boson~\cite{Alonso:2012px,Pich:2013fba,Jenkins:2013zja,Degrande:2012wf,Buchalla:2013rka,Buchalla:2012qq}. Our approach has been to concentrate in the operators necessary for $W_L W_L$ and $\varphi\varphi$ scattering at tree-level and one loop, simplifying the computations by means of the equivalence theorem~\cite{Appelquist}.
Our effective Lagrangian is general enough to include as particular cases the low-energy representation of dilaton models~\cite{Grinstein}, composite Higgs models~\cite{SO(5)}, the old electroweak chiral Lagrangian~\cite{GB,DHD}
and also trivially the Standard Model~\cite{GWS}, and other conceivable models as long as the couplings $\lambda_4\varphi^3$ and $\lambda_4 \varphi^4$ are of order $M_\varphi^2$ and thus small at large energy. We explicitly exclude those models where these Higgs self-couplings take large values~\cite{Siringo:2001hm} that require further analysis. Our $\omega$, $\varphi$ fields are derivatively coupled as befits Goldstone bosons.

We have calculated, in the chiral limit, the unitarized tree-level scattering amplitudes of $\omega\omega\rightarrow\omega\omega$, $\omega\omega\rightarrow\varphi\varphi$ and $\varphi\varphi\rightarrow\varphi\varphi$ processes. Several unitarization methods have been used. And we have also computed the one-loop amplitudes (in the limit $g=g'=0$). We have presented here the elastic $\omega\omega$ amplitude.
Where comparable, our results agree with those of~\cite{Espriu:2012ih}.

We are currently unitarizing the one-loop amplitudes also, but the analysis is highly non-trivial, due to the presence of 8 parameters in the effective Lagrangian (this can be reduced to 7 by rescaling). 
In any case, strong interactions (that is, saturation of unitarity) takes place either when $\beta\neq\alpha^2$, $f\neq v$ or, at one-loop, when the renormalization parameters are non-vanishing. That this happens at the TeV scale is dependent on the parameter choice, but this can be established at the LHC from precise low-energy data.

In the tree-level case, there are different regions in our parameter space. For $\alpha^2=\beta$, $f>v$  strong elastic interactions are expected for $W_LW_L$, and a second, broad scalar structure analogous to the $\sigma$ in nuclear physics possibly appears. We identify a pole at 800 GeV or above in the second Riemann sheet very clearly. However this pole can hardly be considered a resonance  because
it is too broad.

Even if $f\simeq v$ and with small $\lambda_i$, but as long as we allow $\beta>\alpha^2$, we can have strong dynamics resonating between the $W_LW_L$ and $\varphi\varphi$ channels, likewise possibly generating a new scalar pole of the scattering amplitude in the sub-TeV region.

And, finally, the MSM remains as an exceptional case of a light weakly interacting Higgs, which is reached at $\beta=\alpha^2$ and $f=v$. Precision analysis at the LHC  may reveal departures from this minimal scenario~\cite{Gupta:2013zza}.

In that case, new physics in the presence of an electroweak mass gap would very likely imply strong interactions, in elastic $W_L W_L$ and inelastic $\to \varphi\varphi$ scattering.


\begin{theacknowledgments}
AD thanks useful conversations with D. Espriu, M. J. Herrero and J. J. Sanz-Cillero
and the organizers of the second Russian-Spanish Congress for their warm 
hospitality at Saint Petersburg. This work has been supported by the spanish grant FPA2011-27853-C02-01 and by the grant BES-2012-056054 (RLD).

\end{theacknowledgments}

\bibliographystyle{aipproc}   

\end{document}